\documentclass[twocolumn]{revtex4}
\usepackage[dvips]{graphicx}
\usepackage{float,color}

\begin{document}

\title{Surface scattering effect and the stripe order in films of the superfluid $^3$He B phase}

\author{Kazushi Aoyama}

\affiliation{Department of Earth and Space Science, Graduate School of Science, Osaka University, Toyonaka, Osaka 560-0043, Japan 
%${}^2$ The Hakubi Center for Advanced Research, Kyoto University, Kyoto 606-8501, Japan \\
%${}^3$ Department of Physics, Kyoto University, Kyoto 606-8502, Japan 
}

\begin{abstract}
Surface scattering effects in thin films of the superfluid $^3$He B phase have been theoretically investigated, with an emphasis on the stability of the stripe order with spontaneous broken translational symmetry in the film plane and quasiparticle excitations in this spatially inhomogeneous phase. Based on the Ginzburg-Landau theory in the weak coupling limit, we have shown that the stripe order, which was originally discussed for a film with two specular surfaces, can be stable in a film with one specular and one diffusive surfaces which should correspond to superfluid $^3$He on a substrate. It is also found by numerically solving the Eilenberger equation that due to the stripe structure, a midgap state distinct from the surface Andreev bound state emerges and its signature is reflected in the local density of states. 
\end{abstract}
\maketitle

\section{Introduction}
In non-$s$-wave Cooper pairing states, quasiparticle scatterings from system surfaces and nonmagnetic impurities can lead to suppression of the order parameter. One typical example of such anisotropic pairing states would be the spin-triplet $p$-wave one which is realized in superfluid phases of liquid $^3$He. In contrast to superconductors whose properties are closely connected to the underlying crystal symmetries, bulk superfluid $^3$He is a highly clean homogeneous system, so that the Cooper pair can potentially take all the possible symmetries within the spin-triplet $p$-wave sector. In reality, however, only two superfluid phases are realized in the bulk ${}^3$He, B and A phases which are respectively identified with the Balian-Werthamer (BW) and the Anderson-Brinkman-Morel (ABM) pairing states \cite{VW}. The former is a full gapped state appearing in the wide range of the temperature-pressure phase diagram, and the latter with point nodes at both poles of the Fermi sphere is stabilized by the strong-coupling effect only in the high-temperature and high-pressure region. The most interesting question is whether or not pairing states other than the ABM and BW ones may appear in superfluid $^3$He confined in thin slabs, narrow cylinders, and porous media such as aerogels. Since orbital degrees of freedom of the superfluid order parameter are strongly restricted due to the scatterings from such foreign objects, the pure bulk states can be unstable in these geometries. 

In the BW state in a film geometry, surface-normal components of the order parameter are locally suppressed near specular system surfaces. As the film thickness becomes small, these components get smaller even inside the film and the BW state is eventually deformed into the planar pairing state the nodal structure of which is the same as that of the ABM state \cite{film_Nagato, film_Sauls, film_Murakawa}. The relative stability between the planar-distorted BW pairing state and the ABM one depends on the strength of the strong-coupling effect which varies as a function of pressure \cite{Li-Ho}. In experiments, it has been observed that the stability region of the A phase is extended in thin slabs \cite{film_Science, film_PRL}. On the other hand, in aerogels, interconnected silica strands work as nonmagnetic impurities for superfluid $^3$He, giving rise to intriguing phenomena. There, pinning effects of the $l$-vector in the ABM state have been extensively studied \cite{Volovik_aero1, AI_is, AI_anis, Kunimatsu, Bunkov, Volovik_aero2}, and the polar pairing state with a line node along the equator of the Fermi sphere was predicted to appear in uniaxially stretched media \cite{AI_anis}. Recently, the existence of this polar phase has been experimentally confirmed in nematically-ordered aerogels \cite{Dmitriev1, Dmitriev2}. Also in narrow cylinders, the occurrence of the polar state is theoretically predicted \cite{Fetter, Wiman_cylinder}, but so far, any signature of this state has not been observed in experiments \cite{cylinder_Manninen, cylinder_Kotsubo, cylinder_Pekola, cylinder_Saunders, cylinder_Yamaguchi}. The key ingredient common to the above cases is a global anisotropy. It lowers the effective dimensionality of the system favoring the anisotropic pairing states. In such quasi-low dimensional systems, one may naturally expect that the superfluid state is spatially uniform along scattering-free directions, namely, the in-plane and cylinder-axis directions for the slabs and cylinders, respectively. Near the continuous transition from the BW state into the planer or polar states, however, the surface scattering yields periodic spatial structures along these scattering-free directions \cite{Vorontsov_SF, Aoyama_cylinder, Wiman_film}. 

From recent theoretical studies of the surface scattering effect on unconventional superconductors and superfluids, it is becoming clear that the surface-induced gap distortion triggers instability of spatially modulated Cooper pairing states in quasi-low dimensional systems \cite{Vorontsov_SF, Vorontsov_SC, Hachiya, Aoyama_cylinder, Higashitani_filmFFLO1, Higashitani_filmFFLO2, Fogelstrom, Vorontsov_review, Wiman_film}. They are analog of the Fulde-Ferrell-Larkin-Ovchinnikov (FFLO) state \cite{FF, LO} which was originally studied in the context of spin-singlet superconductors in a strong Zeeman field, and are characterized by a nonzero center of mass momentum of the Cooper pair ${\bf Q}$. 
The occurrence of the FFLO-like ${\bf Q}\neq 0$ state in superfluid $^3$He was first pointed out by Vorontsov and Sauls for a film with two specular surfaces \cite{Vorontsov_SF}. Near the BW-planar transition in the film, the BW state exhibits a one-dimensional modulation characterized by ${\bf Q}$ within the film plane. This modulated BW state is called "stripe order". After their prediction, it is shown that the stripe order may be possible also in cylindrical geometry \cite{Aoyama_cylinder}. 
%In principle, a modulated ABM state analogous to the stripe order may appear in some cases. Indeed, it is shown that such a ABM state is favored in a cylinder \cite{Wiman}. 
In narrow cylinders, however, the stripe order can be stable only when the system surface is specular and diffusive along the cylinder axis and rim, respectively. This indicates that the stability of the stripe order is sensitive to whether the surface scattering is specular or diffusive. For the slab geometry, it is not clear whether the stripe order can survive for any scattering conditions or not (quite recently, the strong-coupling effect and surface-roughness effect have been theoretically studied \cite{Wiman_film}, and we will discuss these issues later on). In addition, it is also an interesting question how the midgap states in the $^3$He B phase are modified by the periodic stripe structure.
   
In this paper, we will consider thin films of superfluid $^3$He and investigate (1) surface-roughness effects on the stripe order and (2) quasiparticle excitations in this spatially modulated phase. Throughout this paper, we restrict ourselves to liquid $^3$He at 0 bar which is considered to be in the weak-coupling limit \cite{VW}. It will be shown that the stripe order can exist even in a film with one specular and one diffusive surfaces and that the stripe structure induces a new gapped bound state distinctly different from the conventional surface Andreev bound state.

The remainder of this paper is organized as follows: In Sec.II, we introduce the Ginzburg-Landau (GL) free-energy functional and examine the surface scattering effects on the BW pairing state. The mechanism of the stripe order and its stability against surface roughness are discussed based on the GL theory. In Sec.III, we investigate low-energy quasiparticle excitations in the stripe order by solving the Eilenberger equation to obtain the local density of states (LDOS). There, for simplicity, we have used the spatial profile of the order parameter obtained in Sec.II, i.e., in the GL theory. Finally, we summarize our results in Sec.IV.

%%%%%%%%%%%%%%%%%%%%%%%%               
\section{stripe order in thin films of the superfluid $^3$He B phase} 
In this section, we will discuss the stability of the stripe order in the BW pairing state confined in a quasi-two dimensional system with thickness $D$. The system geometry is shown in Fig.\ref{fig:4}(a). The two-dimensional plane is extending in the $x$-$y$ plane, and quasiparticles are scattered on the system surfaces located at $z=\pm D/2$. Hereafter, three patterns of surface conditions will be considered: (A) both surfaces are specular, (B) the upper surface is specular while the lower one is diffusive, and (C) both surfaces are diffusive. The case (A) corresponds to the ideal superfluid $^3$He film which can be realized by coating slab surfaces with $^4$He \cite{specularity}. The situations (B) and (C) describe superfluid $^3$He on a rough substrate and in a slab without the $^4$He coat, respectively. We will first visit the simplest case (A), which has been already discussed in \cite{Vorontsov_SF}, to clarify the mechanism of the stripe order near the BW-planer transition, and next examine the surface roughness effect. Throughout this section, we ignore the ABM pairing state because it has the same condensation energy as that of the planar pairing state in the weak coupling limit. 

%**********%
\subsection{Ginzburg-Landau theory}
The gap function of the spin-triplet $p$-wave Cooper pairing state is generally given by $\hat{\Delta}(\hat{{\bf p}}, {\bf r}) =  i \,( \sigma_\mu \sigma_y ) \, A_{\mu j}({\bf r})\hat{p}_j$ with Pauli matrices $\sigma_\mu$ $(\mu = x,y,z)$ and complex variables $A_{\mu j}({\bf r})$ which play a role of the order parameter of this system. Properties of liquid $^3$He near the superfluid transition temperature $T_c$ are well described by the GL theory \cite{VW}. As our main focus is on pairing states of a superfluid film with its thickness $D$ less than the dipole length $\sim 12 \, \mu$m, we will neglect the dipole interaction. 
The corresponding functional is obtained as an expansion in the order parameter $A_{\mu j}({\bf r})$ 
\begin{eqnarray}\label{eq:GL}
{\cal F}_{\rm GL}&=&\int_0^{L_x} dx \int_0^{L_y} dy \int_{-D/2}^{D/2} dz \Big( f_{\rm bulk} + f_{\rm grad}\Big), \\
f_{\rm bulk} &=& \alpha A_{\mu i}^\ast A_{\mu i}+\beta_1 |A_{\mu i}A_{\mu i}|^2+\beta_2 (A_{\mu i}^{\ast}A_{\mu i})^2 \nonumber\\
&+& \beta_3 A_{\mu i}^{\ast}A_{\nu i}^{\ast}A_{\mu j}A_{\nu j} + \beta_4 A_{\mu i}^{\ast}A_{\nu i}A_{\nu j}^{\ast}A_{\mu j} \nonumber\\
&+& \beta_5 A_{\mu i}^{\ast}A_{\nu i}A_{\nu j}A_{\mu j}^{\ast} , \nonumber\\
f_{\rm grad}&=& K_1 \big(\nabla_j A^*_{\mu i}\big) \big(\nabla_j A_{\mu i} \big) + K_2 \big( \nabla_j A^*_{\mu i}\big)\big(\nabla_i A_{\mu j}\big) \nonumber\\
&+& K_3 \big( \nabla_i A^*_{\mu i} \big) \big(\nabla_j A_{\mu j} \big) .
\end{eqnarray} 
In the weak coupling limit, the coefficients are given by $\alpha=\frac{1}{3}N_F\ln(T/T_c)$, $-2\beta_1=\beta_2 = \beta_3 = \beta_4= -\beta_5 =2\beta_0$, $\beta_0\equiv 7\zeta(3)N_F/(240\pi^2T^2)$, and $K_1=K_2=K_3=K\equiv \frac{1}{5} N_F (T_c/T)^2 \xi_0^2$ with $N_F$ as density of states per spin on the Fermi surface and $\xi_0 = (v_{\rm F}/2 \pi T_c) \sqrt{7\zeta(3)/12}$ as the superfluid coherence length at $T=0$. Note that this definition of $\xi_0$ is based on Ref.\cite{VW} and is different from that in Ref.\cite{film_Sauls, Vorontsov_SF, Wiman_film} by the factor of $\sqrt{7\zeta(3)/12}=0.837$. 

The effect of the system surface can be incorporated by the boundary condition on the order parameter $A_{\mu i}$. For the specular surface scattering, the surface-normal component of the quasiparticle momentum $\hat{\bf p}$ changes its sign by the mirror reflection at $z= \pm D/2$, i.e., $\hat{\bf p} \rightarrow \underline{\hat{\bf p}} = \hat{\bf p}-2 \hat{\bf n} \, (\hat{\bf n} \cdot \hat{\bf p})$, where $\hat{\bf n}$ denotes a unit vector normal to the surface. Because the gap function should satisfy the equation $\hat{\Delta}(\hat{{\bf p}}, z=\pm D/2)=\hat{\Delta}(\hat{\underline{\bf p}}, z=\pm D/2)$ at the surfaces, the surface-normal component of the order parameter must vanish at the system surface while parallel components should be unchanged. Thus, the boundary condition on $A_{\mu i}$ at the specular surface reads   
\begin{eqnarray}\label{eq:boundary_sp} 
&& A_{\mu z}(x,y, \, z=\pm D/2 )=0, \nonumber\\
&& \nabla_z A_{\mu i}(x,y, \, z=\pm D/2 )=0  \quad (i \neq z).
\end{eqnarray}
%This guarantees that there are no net surface terms when we integrate by parts the gradient terms.
When the surface is sufficiently rough such that quasiparticles are randomly scattered independent of the incident direction, the boundary condition Eq.(\ref{eq:boundary_sp}) is replaced with the diffusive one, i.e., $A_{\mu i}=0$ for any $\mu$ and $i$.

Now, we turn to the instability of the stripe order with ${\bf Q}\neq 0$ in the BW state. In the BW state in the slab geometry, the order parameter is usually assumed to be uniform in the two-dimensional plane ($x$-$y$ plane), and its basic form is given by $A^{(B)}_{\mu i}(z)=a_{xx}(z)\hat{x}_\mu\hat{x}_{i}+a_{yy}(z)\hat{y}_\mu\hat{y}_{i}+a_{zz}(z)\hat{z}_\mu\hat{z}_{i}$. Because of the boundary condition Eq.(\ref{eq:boundary_sp}), $a_{zz}(z)$ vanishes at $z=\pm D/2$. It has been known that due to the surface-induced gap suppression, the BW state is gradually deformed into the planar pairing state $A^{({\rm planar})}_{\mu i}=\Delta_1(\hat{x}_\mu\hat{x}_{i}+\hat{y}_\mu\hat{y}_{i})$ as film thickness $D$ becomes small \cite{film_Nagato, film_Sauls}. Below, we will see that near the BW-planar transition $T_{BP}$, the BW state lowers the energy by introducing the spatial modulation ${\bf Q}$ in the $x$-$y$ plane and extends its stability region to higher temperatures. 
For brevity, we take into account spatial variations in the order parameter only along the $x$ direction, assuming that the modulation, namely, periodic stripe structure is introduced in the $x$ direction (see Fig.\ref{fig:4}(a)). As $x$ components of the order parameter are generally relevant to spatial variations in the $x$ direction, we start from the BW state of the form  
\begin{eqnarray}\label{eq:BW_op}
&& A_{\mu i}(x,z) \\
&& = \left( \begin{array}{ccc}
A_{xx}(x,z) & 0 & A_{xz}(x,z) \\
0 & A_{yy}(x,z) & 0 \\
A_{zx}(x,z) & 0 & A_{zz}(x,z) 
\end{array} \right). 
\end{eqnarray}

By expanding $A_{\mu i}(x,z)$ in a Fourier series with respect to $x$ and picking up relevant leading order terms, we have the following form of the order parameter which can be continuously deformed into $A^{({\rm planar})}_{\mu i}$:
\begin{eqnarray}\label{eq:BW_op_Q}
&& A_{\mu i}(x,z) \\
&& = \left( \begin{array}{ccc}
a_{xx}(z) & 0 & 0 \\
0 & a_{yy}(z) & 0 \\
a_{zx}(z)\sin(Qx) & 0 & a_{zz}(z)\cos(Qx) 
\end{array} \right), \nonumber
\end{eqnarray}
where $Q=2\pi n/L_x$ is the wave vector characterizing the spatial modulation of the order parameter in the $x$ direction, and $A_{xz}$ has been dropped because its contribution is known to be negligibly small \cite{Vorontsov_SF}. As we will see below, not only the off-diagonal component itself $A_{zx}$ but also the phase difference of $\pi/2$ between sine waves in $A_{zx}$ and $A_{zz}$ is important for the occurrence of the stripe order. 

\subsection{Mechanism of the stripe order}   

\begin{figure}[t]
\includegraphics[scale=0.55]{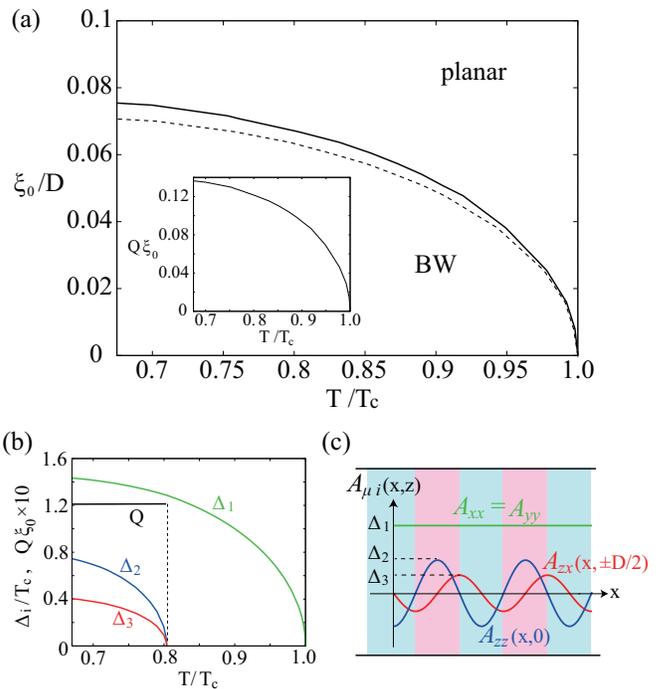}
\caption{(color online) (a) The BW-planar transition curves $T_{BP}(D)$ in a film with thickness $D$ and specular surfaces. $T_{BP}$ for the BW state with a one-dimensional spatial modulation in the film plane (solid curve) is higher than that for the uniform BW state (black dashed one). Inset shows the evolution of the characteristic wave number $Q$ along the $T_{BP}(D)$ curve. (b) Temperature dependences of $\Delta_1$ (green), $\Delta_2$ (blue), $\Delta_3$ (red), and $Q$ (black) at $D/\xi_0=15$. (c) The spatial profile of the order parameter along the $x$-axis in the striped BW state with ${\bf Q}\parallel \hat{x}$.  
\label{fig:1}}
\end{figure}

The mechanism of the stripe order in the cylindrical geometry has been already addressed in Ref.\cite{Aoyama_cylinder}. The origin of the stripe order in film superfulid $^3$He can be understood in a similar way. We first consider the case with specular surface scattering and take the following trial state satisfying Eq.(\ref{eq:boundary_sp}),
\begin{eqnarray}\label{eq:OP_trial}
&& a_{xx}(z)=a_{yy}(z)=\Delta_1 ,    \nonumber\\
&& a_{z z}(z) = \Delta_2 \cos\big(\frac{\pi z}{D} \big),  \nonumber\\
&& a_{z x}(z) = \Delta_3 \sin\big(\frac{\pi z}{D} \big) .
\end{eqnarray}
Translational symmetry breaking in the BW state is signaled by finite values of $Q$. Inserting the expression Eq.(\ref{eq:OP_trial}) into Eq.(\ref{eq:GL}) and integrating over $z$ and $x$, we obtain
\begin{eqnarray}\label{eq:GL_trial}
{\cal F}_{\rm GL}&=& D L_x L_y \Big[ \alpha \sum_i a_i \Delta_i^2 +\beta_0 \sum_{i\leq j} b_{ij} \Delta_i^2 \Delta_j^2 + K f_g  \Big], \nonumber\\
f_g &=& \Delta_2^2 \Big(\frac{3}{4} \frac{\pi^2}{D^2}+ \frac{1}{4} Q^2 \Big) +\Delta_3^2 \Big( \frac{1}{4}\frac{\pi^2}{D^2} +\frac{3}{4}Q^2 \Big) \nonumber\\
&-& \Delta_2 \Delta_3 \, Q \, \frac{\pi}{D} \nonumber\\
&=& C_Q \Big( Q-\frac{ \pi \Delta_2 \Delta_3}{2 C_Q \, D} \Big)^2 +\frac{\pi^2 \Delta_2^2}{4 D^2}\Big( 3  + \frac{\Delta_3^2}{\Delta_2^2} -\frac{\Delta_3^2}{C_Q}\Big) 
\end{eqnarray}
with $C_Q= ( \Delta_2^2 + 3 \Delta_3^2 )/4$. The coefficients are calculated as $a_1=2$, $a_2=a_3=1/4$, $b_{11}=8$, $b_{12}=1$, $b_{13}=2$, $b_{22}=b_{33}=27/64$, and $b_{23}=3/32$.
It should be emphasized here that the gradient term linear in $Q$ shows up. By minimizing $f_g$ with respect to $Q$, we obtain 
\begin{equation}
Q= \pi \Delta_2 \Delta_3/(2 C_Q \, D )
\end{equation}
which is nonzero as long as $\Delta_2$ and $\Delta_3$ are nonzero. The gradient energy for the optimal $Q$ is lower than that for $Q=0$ by $ \pi^2 \Delta_2^2\Delta_3^2 /( 4 C_Q \, D^2 )$. Thus, the system tends to introduce the modulation $Q$ to lower $f_g$ leading to the stripe order. This situation is in sharp contrast to conventional FFLO states for which a $Q$-linear gradient term does not exist and nonzero $Q$ appears just because the coefficient of the $Q^2$ term becomes negative at very low temperatures for strong Zeeman fields \cite{Adachi}. The present system is, however, rather similar to non-centrosymmetric superconductors (NCS) with Rashba spin-orbit coupling \cite{Rashba, Sigrist_book} in a magnetic field where the so-called helical phase with a field-induced phase modulation $e^{i {\bf Q}\cdot {\bf r}}$ is believed to be realized \cite{Dimitrova, Samokhin, Kaur, AS}. In Rashba-type NCS, broken inversion symmetry allows a $Q$-linear gradient term coupled with the field, and as a result, the phase modulation exists even at the superconducting transition temperature in the magnetic field. From the analogy to NCS, it is inferred that the modulated BW state would emerge from high temperatures near $T_c$.   

Figure\ref{fig:1}(a) shows the $T_{BP}(D)$ curves for the BW states of the form Eq.(\ref{eq:OP_trial}) with (solid curve) and without (dashed one) a one-dimensional modulation in the two-dimensional plane. The $T_{BP}$ transition temperature is higher for $Q \neq 0$ than that for $Q=0$, which implies that the superfluid state with broken translational symmetry is stabilized as the lowest energy BW state. It is striking that the stripe order with $Q \neq 0$ appears from relatively high temperatures near $T_c$, while the conventional FFLO state is realized only in the low temperature region. In the superfluid $^3$He film, the $Q$-linear term coupled with $\Delta_3$ (amplitude of $A_{zx}$) yields the $Q \neq 0$ pairing state, and thus, the internal degrees of freedom of the order parameter play an essential role for the striped superfluid phase. We also note that the $Q$-linear term exists only when we have the phase difference of $\pi/2$ between sine waves in $A_{zx}$ and $A_{zz}$. 

In Fig.\ref{fig:1}(b), one can see that $\Delta_2$ and $\Delta_3$ grow up with the $\sqrt{T_{BP}-T}$ dependence suggestive of the second order BW-planar transition. The inset of Fig.\ref{fig:1}(a) shows the evolution of the modulation $Q$ along the $T_{BP}$ transition curve. $Q$ develops with decreasing film thickness. 
The result obtained here is valid only when the trial state Eq.(\ref{eq:OP_trial}) well describes the exact spatial profile of the order parameter. 
In particular, for rough surfaces, $z$-dependence of $A_{\mu i}$ is not trivial. In the next subsection, we will determine the stability region of the stripe order by numerically solving GL equations.

%**********%
\subsection{Surface-roughness effect on the stripe order}

\begin{figure}[t]
\includegraphics[scale=0.5]{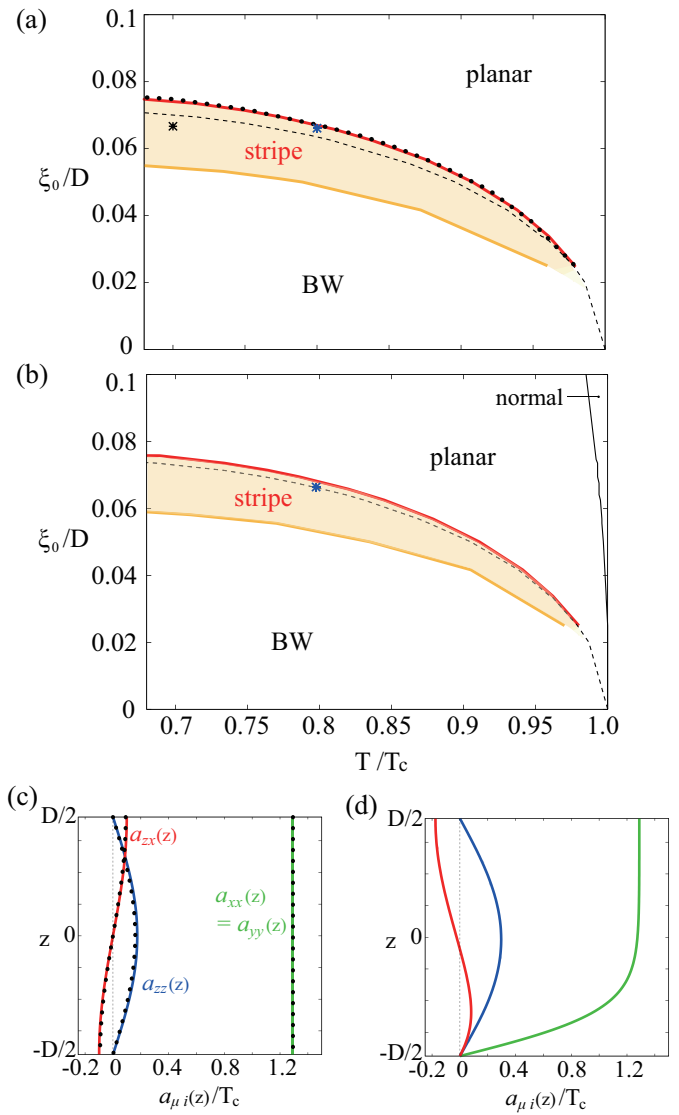}
\caption{(color online) The superfluid $^3$He B phase confined in film geometries with thickness $D$. The T-D$^{-1}$ phase diagram for two specular (one specular and one diffusive) surfaces is shown in (a) [(b)]. The spatial profiles of the order parameter components along $z$-axis at the point $D/\xi_0=15$ and $T/T_c=0.8$, which is denoted by the blue colored symbol * in (a) [(b)], are shown in (c) [(d)]. Green, blue, and red curves in (c) and (d) denote the spatial variations of $a_{xx}=a_{yy}$, $a_{zz}$, and $a_{zx}$, respectively. In (a)-(d), the solid and dashed curves are obtained by numerically solving GL equations, whereas the black dotted ones in (a) and (c) are the results obtained by the variational method with the use of the trial functions Eq.(\ref{eq:OP_trial}). 
%The upper (red solid curve) and lower (orange solid one) boundaries of the stripe order in (a) and (b), the transitions between the uniform BW and the planar states (black dashed one), the planar-normal transition in (b) (black solid one), and the $z$-dependences of $a_{\mu i}$ in (c) and (d) (colored solid ones) are obtained by numerically solving GL equations, whereas the black dots in (a) and (c) are results obtained with the use of Eq.(\ref{eq:OP_trial}). 
\label{fig:2}}
\end{figure}

Here, we consider the three patterns of surface conditions (A) specular-specular, (B) specular-diffusive, and (C) diffusive-diffusive. To obtain the $T-D^{-1}$ phase diagram in each case, we will solve GL equations $\delta {\cal F}_{\rm GL}/\delta A^*_{\mu i}=0$ numerically under corresponding boundary conditions for various values of $Q$ at fixed points of $T$ and $D$, and determine the lowest energy state. For the specular surface, the boundary condition Eq.(\ref{eq:boundary_sp}) is imposed, while for the diffussive one, $A_{\mu i}=0$ is used.
 
Figures \ref{fig:2}(a) and (b) show the numerically obtained $T-D^{-1}$ phase diagrams of the superfluid ${}^3$He B-phase film with surfaces of type (A) and (B), respectively. In the region sandwiched by red and orange curves, the striped BW state with the modulation $Q\neq 0$ is more stable than the uniform one with $Q = 0$. The upper boundary is determined from the single mode instability with the use of Eq.(\ref{eq:BW_op_Q}), and the lower boundary denotes the transition from the uniform BW state into the inhomogeneous one consisting of two domains with opposite signs of $A_{zz}$. In obtaining the lower boundary, we have used the general form of $A_{\mu i}$, Eq.(\ref{eq:BW_op}). Comparing Figs.\ref{fig:2}(a) and (b), one can see that the stability region of the stripe order shrinks as the surface scattering becomes more diffusive. 

Figure\ref{fig:2}(c) shows $z$-dependences of $a_{\mu i}$ near the upper boundary of the stripe-order stability region in a film with two specular surfaces, where solid and dotted curves denote the results obtained by numerically solving GL equations and by the variational method with the use of the trial state Eq.(\ref{eq:OP_trial}), respectively. One can see that the trial state Eq.(\ref{eq:OP_trial}) well describes the spatial profiles of $a_{z i}(z)$. Correspondingly, the stripe-planar transition curve obtained by solving GL equations is in quantitative agreement with the one calculated with the variational method based on Eq.(\ref{eq:OP_trial}). 
On the other hand, in the case (B), as is shown in Fig.\ref{fig:2}(d), $a_{\mu i}(z)$ are squashed to zero at the lower diffusive surface, and as a result, $a_{\mu i}(z)$ exhibit large spatial variations near the diffusive surface. Although such large spatial variations cost much gradient energy, the stripe order is still stable as the $Q$-linear gradient term can lower the {\it net} gradient energy like in the case with two specular surfaces. 

In the case (C) where both surfaces are diffusive, $a_{\mu i}(z)$ must be zero at the both upper and lower surfaces, which results in a large spatial variation in $a_{z x}(z)$ in the striped BW state with $Q \neq 0$. By using the spatial profiles of $a_{\mu i}(z)$ compatible with the diffusive condition
\begin{eqnarray}\label{eq:OP_trial_df}
&& a_{xx}(z)=a_{yy}(z)=\Delta_1 \cos\big(\frac{\pi z}{D} \big),    \nonumber\\
&& a_{z z}(z) = \Delta_2 \cos\big(\frac{\pi z}{D} \big),  \nonumber\\
&& a_{z x}(z) = \Delta_3 \sin\big(\frac{2\pi z}{D} \big),
\end{eqnarray}
we can evaluate the gradient energy as  
\begin{eqnarray}\label{eq:GL_trial_df}
f_g &=& \Delta_2^2 \Big(\frac{3}{4} \frac{\pi^2}{D^2}+ \frac{1}{4} Q^2 \Big) +\Delta_3^2 \Big( \frac{\pi^2}{D^2} +\frac{3}{4}Q^2 \Big) \nonumber\\
&-& \frac{8}{3\pi}\Delta_2 \Delta_3 \, Q \, \frac{\pi}{D} + \Delta_1^2 \frac{\pi^2}{D^2}.
\end{eqnarray}
Comparing Eqs.(\ref{eq:GL_trial}) and (\ref{eq:GL_trial_df}), one can see that $Q$-relevant terms are almost unchanged, while the associated energy cost for $a_{z x}(z)$, which corresponds to $\frac{\pi^2}{D^2}\Delta_3^2$ in Eq.(\ref{eq:GL_trial_df}), is much enhanced. Because the large spatial variation in $a_{z x}(z)$ lowers the net energy gain, the stripe order becomes less stable. In the variational method using Eq.(\ref{eq:OP_trial_df}), $T_{BP}(D)$ for the $Q \neq 0$ state is higher than the one for the $Q = 0$ state only by less than $1 \%$ of $T_{BP}(D)$ for $\xi_0/D < 0.075$.
Such a result is also obtained from exact numerical solutions of the GL equations. Thus, the uniform and striped BW states are almost degenerate near $T_{BP}(D)$. This suggests that the stability region of the stripe phase is restricted in the vicinity of the $T_{BP}(D)$ curve. In fact, numerical calculations for determining the lower phase boundary show that the uniform BW phase extends up to near $T_{BP}(D)$ and the striped BW phase is possible only in a very narrow region below $T_{BP}(D)$. It is, however, difficult to accurately determine the lower phase boundary because near $T_{BP}(D)$ the free energy difference between the two states is very small compared with numerical errors. The phase diagram in the case (C) is therefore not shown in Fig.\ref{fig:2}. 

%In this case, the stripe order is not stable any more, and if it exists, it should be restricted in the vicinity of the $T_{BP}(D)$ curve. 
%In the recent theoretical work taking account of the strong-coupling effect \cite{Wiman_film}, the same conclusion has been reported for a film with two diffusive surfaces.

The above our results are obtained in the weak coupling limit without the strong-coupling effect which favors the ABM pairing state in the bulk. In general, when the strong-coupling corrections are incorporated, the stability region of the stripe order may be modified. Recent theoretical work has shown that even in the presence of the strong-coupling contributions, the stripe order can survive for moderately diffusive surfaces with its stability region being suppressed at high temperatures, while it exists only in the vicinity of $T=0$ for maximally diffusive surface scatterings \cite{Wiman_film}. Although in Ref.\cite{Wiman_film}, it is assumed that the strong-coupling contributions in a slab can be evaluated from the corresponding bulk values, they might be quite different from those in the bulk like in the case of superfluid $^3$He in globally anisotropic aerogels \cite{Ikeda_aniso-sc}, but this issue is beyond the scope of this work.

%%%%%%%%%%%%%%%%%%%%%%%%
\section{Quasiparticle excitations in the stripe order}
In this section, we will investigate quasiparticle excitations in the stripe order. As our focus is on how the midgap state is affected by the stripe structure in the pair potential $\hat{\Delta}(\hat{{\bf p}}, {\bf r}) =  i \,( \sigma_\mu \sigma_y ) \, A_{\mu j}({\bf r})\hat{p}_j$, we will consider the typical case with two specular surfaces. For the specular scattering, we have seen in the previous section that the simplified form of the order parameter Eq.(\ref{eq:OP_trial}) well describes superfluid properties near the stripe-planar transition, so that we could expect that Eq.(\ref{eq:OP_trial}) should work also for examining quasiparticle excitations in the stripe order. Below, we will solve the Eilenberger equation with the use of Eqs.(\ref{eq:BW_op_Q}) and (\ref{eq:OP_trial}) to obtain the angle-resolved LDOS which gives detailed informations on the low-energy excitations.

%**********%
\subsection{quasiclassical theory and numerical methods}
The quasiclassical spinful Eilenberger equation has been extensively used for studies of superfluid $^3$He. In general, it should be solved self-consistently in combination with the superfluid gap equation. As we have mentioned above, however, in this study we will use the fixed form of the gap function Eq.(\ref{eq:OP_trial}) instead of solving the gap equation. This procedure should be valid for qualitative discussion on the low-energy excitations near $T_c$ because in this high temperature regime, the gap equation is reduced to the GL equation whose solution is well approximated by Eq.(\ref{eq:OP_trial}).     

It has been known that the Eilenberger equation can be easily solved by introducing so-called Riccati amplitudes $\hat{a}$ and $\hat{b}$ which are $2\times 2$ matrices obeying the Riccati equations
\begin{eqnarray}\label{eq:riccati}
 && {\bf v}_{\rm F}\cdot \nabla \hat{a} + 2\varepsilon_n \hat{a}+ \hat{a} \hat{\Delta}^\dagger\hat{a}-\hat{\Delta}=0 , \nonumber\\
 && {\bf v}_{\rm F}\cdot \nabla \hat{b} - 2\varepsilon_n \hat{b}- \hat{b} \hat{\Delta}\hat{b}+\hat{\Delta}^\dagger=0 
\end{eqnarray}
with Matsubara frequency $\varepsilon_n=(2n+1)\pi T$ \cite{film_Sauls}. The angle-resolved LDOS $N({\bf r},E,\hat{\bf p})$ can be expressed in terms of the Riccati amplitude $\hat{a}$ and $\hat{b}$ as
\begin{equation}\label{eq:dos}
N({\bf r},E,\hat{\bf p})= N_F \, {\rm Re}\Big\{ \frac{1}{2}{\rm tr}\big[ (1+\hat{a}\hat{b})^{-1}(1-\hat{a}\hat{b}) \big]\Big| _{i\varepsilon_n\rightarrow E+i\eta} \Big\},
\end{equation}
where $\eta$ is a positive infinitesimal constant. The amplitudes $\hat{a}| _{i\varepsilon_n\rightarrow E+i\eta}$ and $\hat{b}| _{i\varepsilon_n\rightarrow E+i\eta}$ can be obtained by solving Eq.(\ref{eq:riccati}) with the replacement $\varepsilon_n \rightarrow \eta-iE$ for the known pair potential. In this paper, the positive constant is chosen to be $\eta=0.01 T_c$. 

Although the numerical integration of the differential equation (\ref{eq:riccati}) requires initial values of $\hat{a}$ and $\hat{b}$, we do not know the initial values anywhere in the film. Thus, we start from initial guess for $\hat{a}$ and $\hat{b}$ and numerically integrate along a classical trajectory until convergence is reached \cite{film_Sauls, Nagai_meso}. The concrete procedure is as follows. Let ${\bf x}^N_{\rm fw}$ (${\bf x}^N_{\rm bk}$) be a point which the classical trajectory starting from ${\bf x}$ along $\hat{\bf p}$ (-$\hat{\bf p}$) reaches after $N$ times reflections at the surfaces. The Riccati amplitude $\hat{a}$ ($\hat{b}$) at a position ${\bf x}$ for momentum $\hat{\bf p}$ are obtained by numerically integrating Eq.(\ref{eq:riccati}) forward (backward) along the classical trajectory with an arbitrary initial value at ${\bf x}^N_{\rm bk}$ (${\bf x}^N_{\rm fw}$). Note that $N$ must be sufficiently large such that the total length of the trajectory is much longer than the superfluid coherence length $\xi_0$ and the obtained Riccati amplitudes do not depend on initial values. In this work, we take $N=50$ and parametrize the momentum direction as $\hat{\bf p}=(\sin\theta \, \cos\phi, \sin\theta \, \sin\phi, \cos\theta )$ (see Fig.\ref{fig:4}(a)).

\begin{figure}[t]
\includegraphics[scale=0.8]{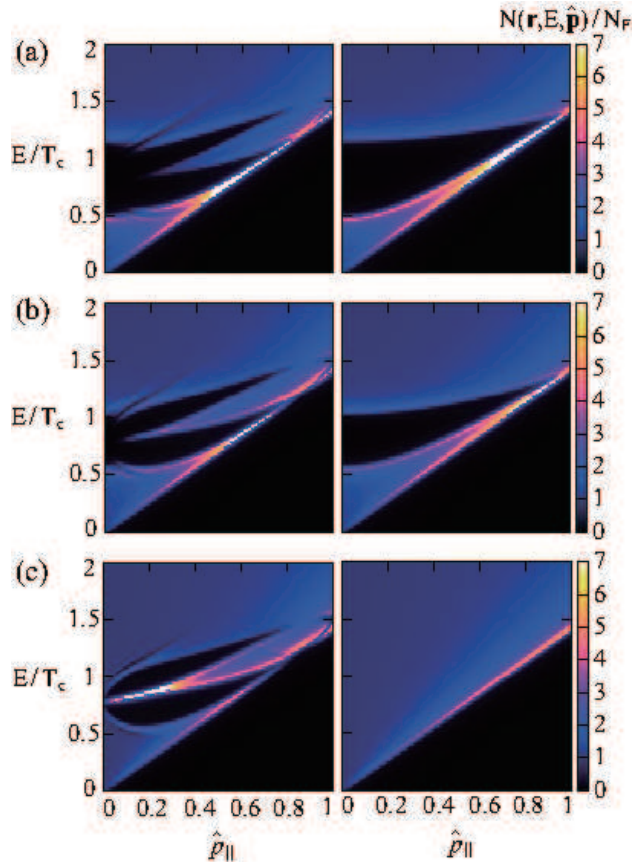}
\caption{(color online) Angle-resolved local density of states (LDOS) in the stripe order at $D/\xi_0=15$ and $T/T_c=0.7$ [the black colored symbol * in Fig.\ref{fig:2} (a)] on the surface of the superfluid film with two specular surfaces. (a), (b), and (c) are results at the positions A ($x/\xi_0=0$), B ($x/\xi_0=\pi/(4Q)$), and C ($x/\xi_0=\pi/(2Q)$) in Fig.\ref{fig:4} (a), respectively. Left (right) panels show LDOS's for the quasiparticle trajectory across (along) the stripe, namely, $\phi=0$ ($\phi=\pi/2$). \label{fig:3}}
\end{figure}

\subsection{local density of states and midgap states} 
\begin{figure}[t]
\includegraphics[scale=0.42]{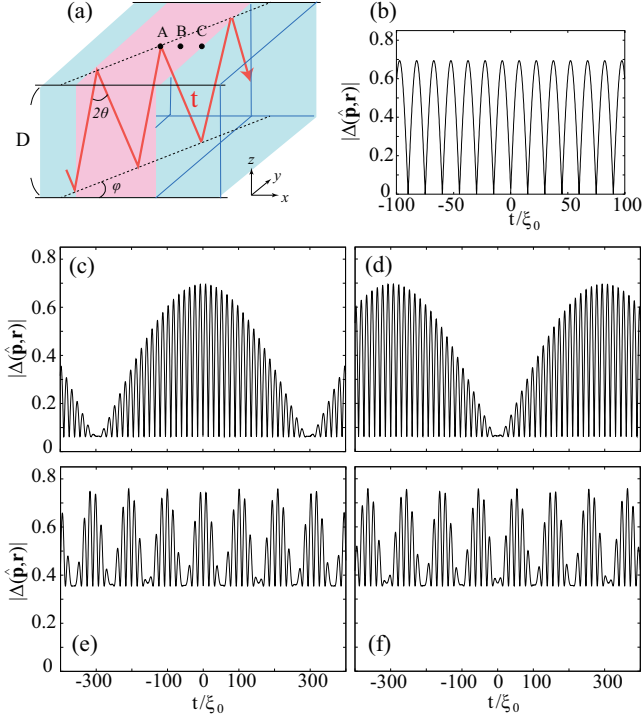}
\caption{(color online) (a) System geometry and a quasiparticle trajectory, where A ($x/\xi_0=0$), C ($x/\xi_0=\pi/(2Q)$), and B ($x/\xi_0=\pi/(4Q)$) on the film surface represent the domain center, the domain boundary, and the middle point between them, respectively. (b) The amplitude of the superfluid gap $|\Delta(\hat{\bf p},{\bf r})|$ at $D/\xi_0=15$ and $T/T_c=0.7$ along the quasiparticle trajectory which passes through the position A with the azimuthal angle $\theta=0.014 \pi$ ($\hat{p}_\parallel=0.043$) and the polar angle $\phi=\pi/2$ (parallel to the stripe). The origin $t=0$ corresponds to the position A. (c)-(f) The superfluid gap $|\Delta(\hat{\bf p},{\bf r})|$ at $D/\xi_0=15$ and $T/T_c=0.7$ along the trajectory running across the stripe with the polar angle $\phi=0$. The origins $t=0$ in left [(c) and (e)] and right [(d) and (f)] panels correspond to the positions A and C, respectively. The azimuthal direction of the trajectory for (c) and (d) [(e) and (f)] is characterized by $\hat{p}_\parallel=0.043$ ($0.25$).  \label{fig:4}}
\end{figure}

Figure \ref{fig:3} shows the angle-resolved LDOS $N({\bf r},E,\hat{\bf p})$ on the surface of the film with the thickness $D/\xi_0=15$ at three points along the modulation (A, B, and C in Fig.\ref{fig:4}(a)), where $\hat{p}_\parallel = \sin\theta$. 
We first consider the quasiparticle trajectory perpendicular to the modulation ($\phi=\pi/2$), or equivalently, parallel to the striped domain structure. Here, at the domain center (the point A in Fig.\ref{fig:4}(a)), $A_{zz}$ is nonzero except just at the surfaces, while at the domain boundary (the point C in Fig.\ref{fig:4}(a)), $A_{zz}$ vanishes everywhere within the domain-wall plane. For $\phi=\pi/2$, the quasiparticle excitations should be essentially the same as those for the uniformly deformed BW state without the modulation. As one can see in the right panel of Fig.\ref{fig:3}(a), $N({\bf r},E,\hat{\bf p})$ at the domain center shows two branches in its low-energy part, which is qualitatively consistent with the result obtained by self-consistently solving the Eilenberger and the gap equations \cite{Mizushima_3HeB}. This suggests that our theoretical approach using the fixed form of the pair potential is a good approximation for the qualitative discussion on the quasiparticle excitations. The upper branch originates from the surface Andreev bound state gapped by the overlap of the wave functions at the two surfaces \cite{Mizushima_3HeB, Tsutsumi_3HeAB}, whereas the lower one corresponds to the gapless excitations due to the surface-induced suppression of $A_{zz}$. As the domain boundary is approached, the energy gap between the Andreev bound state and the bulk continuum becomes smaller because the surface-normal component of the order parameter $A_{zz}$ gradually decreases. As one can see in the right panel of Fig.\ref{fig:3}(c), the gap is completely closed at the domain boundary where $A_{zz}$ vanishes. 

In contrast to the conventional behavior for $\phi=\pi/2$, an additional midgap state shows up for the quasiparticle trajectory running across the stripe ($\phi=0$). In the left panels of Fig.\ref{fig:3}, one can find an additional branch between the Andreev bound state and the bulk continuum, and the end point of this branch approaches $\hat{p}_\parallel=0$ as one goes from the domain center toward the domain boundary. To understand the origin of this new bound state, we will examine the spatial profile of the pair potential along the trajectory \cite{Ichioka_Skyrmion}.

The spatial variation in the amplitude of the pair potential $|\Delta(\hat{\bf p},{\bf r})|=\sqrt{{\rm tr}[ \hat{\Delta}(\hat{\bf p}, {\bf r})\hat{\Delta}^\dagger(\hat{\bf p}, {\bf r}) ]/2}$ along the quasiparticle trajectory is shown in Fig.\ref{fig:4}, where $t=0$ corresponds to a point on the surface (A or C in Fig.\ref{fig:4}(a)). For the trajectory slightly tilted from the surface normal, $|\Delta(\hat{\bf p},{\bf r})|$ is very small at the surfaces [Figs.\ref{fig:4}(b)-(d)], while for a highly tilted one, $|\Delta(\hat{\bf p},{\bf r})|$ is relatively large at the surfaces [Figs.\ref{fig:4}(e) and (f)]. In the case of $\phi=\pi/2$ (parallel to the stripe), $|\Delta(\hat{\bf p},{\bf r})|$ exhibits a simple periodic behavior associated with multiple surface scatterings, which can be seen in Fig.\ref{fig:4}(b). By contrast, in the case of $\phi=0$ (perpendicular to the stripe), an additional longer-period oscillation occurs, as is shown in Figs.\ref{fig:4}(c)-(f). This large oscillation is associated with the one-dimensional modulation of the stripe order. In Figs.\ref{fig:4}(c) and (d), one can see that near $t=0$, the large oscillation works as a confinement potential for the quasiparticles at the domain boundary (Fig.\ref{fig:4}(d)), while it works as a energy barrier at the domain center (Fig.\ref{fig:4}(c)). That's the reason why the bound state exists at the domain boundary (see the left panel of Fig.\ref{fig:3}(c)), while it does not near $\hat{p}_\parallel=0$ at the domain center (see the left panel of Fig.\ref{fig:3}(a)). 
Comparing Figs.\ref{fig:4}(c-d) with (e-f), one finds that as the trajectory is tilted from the surface normal, namely, $\hat{p}_\parallel$ increases, the amplitude of the large oscillation becomes small. This indicates that such a reduction of the energy barrier should lead to the overlap of the wave functions localized at each valley in $|\Delta(\hat{\bf p},{\bf r})|$, resulting in the midgap state appearing from a finite value of $\hat{p}_\parallel$ at the domain center.  

\begin{figure}[t]
\includegraphics[scale=0.5]{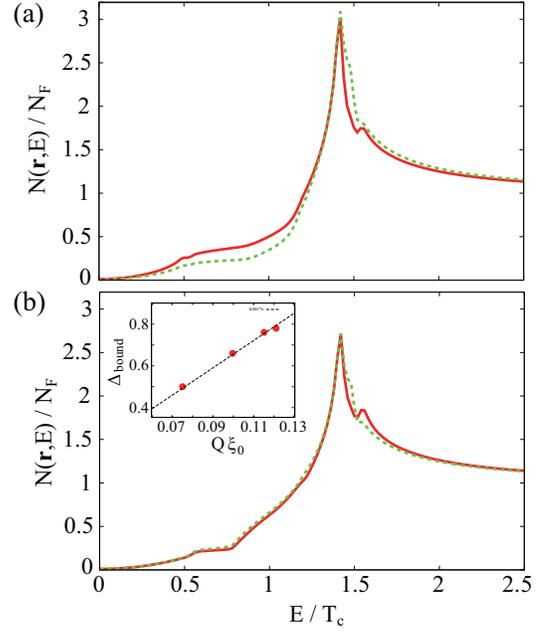}
\caption{(color online) LDOS in the stripe order at $T/T_c=0.8$ for the specular-surface film with the thickness $D/\xi_0=15$. (a) and (b) are results at the domain center A ($x/\xi_0=0$) and the domain boundary C ($x/\xi_0=\pi/(2Q)$), respectively. Red solid (green dashed) curves denote LDOS's at the surface (center) of the film. In the inset of (b), the bound state energy gap $\Delta_{\rm bound}$ is plotted as a function of $Q$; from right to left, data points correspond to $D/\xi_0=15, \, 15.7, \, 18$ at $T/T_c=0.7$, and $D/\xi_0 = 24$ at $T/T_c=0.9$. \label{fig:5}}
\end{figure}

Figures \ref{fig:5}(a) and (b) show the angle-averaged LDOS's at the domain center and boundary for the thickness $D/\xi_0=15$, respectively. The overall feature of the LDOS is almost unchanged along the $z$ axis for a fixed $x$, while the low-energy behavior depends on $x$. As one can see in Fig.\ref{fig:5}(a), the LDOS at the domain center with nonzero $A_{zz}$ exhibits a linear behavior near $E=0$ due to the contribution of the surface states extending from the upper and lower surfaces, which is consistent with the previous result for thin films \cite{Tsutsumi_3HeAB}. On the other hand, one can see in Fig.\ref{fig:5}(b) that the spectrum at the domain boundary has a kink near $E\sim 0.75$. This kink originates from the bound state associated with the modulation of the stripe order. Actually, the kink position coincides with the bound state energy at $\hat{p}_\parallel=0$ $\Delta_{\rm bound} \sim 0.75$ (see the left panel of Fig.\ref{fig:3}(c)). Now, the question is what determines the bound state energy gap $\Delta_{\rm bound}$.

Since the bound state is formed due to the confinement potential shown in Fig.\ref{fig:4}(d), it is natural to expect that $\Delta_{\rm bound}$ is related to the shape of the potential which is characterized by the period of the stripe structure $Q$. In the inset of Fig.\ref{fig:5}(b), $\Delta_{\rm bound}$ is plotted as a function of $Q$ for various film thicknesses. $\Delta_{\rm bound}$ is well scaled by $Q$, suggesting that the bound state energy gap $\Delta_{\rm bound}$ is closely connected to the period of the stripe.

%%%%%%%%%%%%%%%%%%%%%%%%
\section{Conclusion}
In this paper, we have examined the surface-roughness effect on the stability of the stripe order in thin films of the superfluid $^3$He B phase based on the Ginzburg-Landau theory in the weak coupling limit, and also investigated the quasiparticle excitations in this striped superfluid phase by solving the Eilenberger equation.  
Although the occurrence of the stripe order was originally pointed out for a film with two specular surfaces, it is found that the stripe order survives even in a film with one specular and one diffusive surfaces which should correspond to superfluid $^3$He on a substrate. Our numerical results on the angle-resolved local density of states (LDOS) show that a new bound state distinct from the surface Andreev bound state appears for classical trajectories running across the stripe. This unconventional bound state originates from the one dimensional modulation of the order parameter, namely, the stripe structure, and is reflected as a kink in the angle-averaged LDOS.

%%%%%%%%%%%%%%%%%%%%%%%%
\section{Acknowledgement}
The author is grateful to T. Mizushima, K. Machida, and Y. Tsutsumi for useful discussions. This work is supported by a Grant-in-Aid for Scientific Research (Grant No. 25800194).

\end{document}